# Tunable van Hove Singularities and Correlated States in Twisted Trilayer Graphene

Yanmeng Shi[1,2,†], Shuigang Xu[2,†], Mohammed M. Al Ezzi[3,†], Nilanthy Balakrishnan[2,†], Aitor Garcia-Ruiz[1,2], Bonnie Tsim[1,2], Ciaran Mullan[1], Julien Barrier[1,2], Na Xin[2], Benjamin A. Piot[4], Takashi Taniguchi[5], Kenji Watanabe[5], Alexandra Carvalho[3], Artem Mishchenko[1,2], A. K. Geim[1,2], Vladimir I. Fal'ko[1,2,6], Shaffique Adam[3,7], Antonio Helio Castro Neto[3], Kostya S. Novoselov[1,2,3,8,*]

[1]*Department of Physics and Astronomy, University of Manchester, Manchester M13 9PL, UK.*
[2]*National Graphene Institute, University of Manchester, Manchester M13 9PL, UK.*
[3]*Centre for Advanced 2D Materials, National University of Singapore, 117546, Singapore*
[4]*Laboratoire National des Champs Magnétiques Intenses, LNCMI-CNRS-UGA-UPS-INSA-EMFL, 25 avenue des Martyrs, 38042 Grenoble, France.*
[5]*National Institute for Materials Science, 1-1 Namiki, Tsukuba, 305-0044, Japan.*
[6]*Henry Royce Institute for Advanced Materials, Manchester, M13 9PL, UK.*
[7]*Yale-NUS College, 16 College Avenue West, 138527 Singapore.*
[8]*Chongqing 2D Materials Institute, Liangjiang New Area, Chongqing 400714, China*
[†]*These authors contributed equally*
*\*e-mail: kostya@manchester.ac.uk*

**Abstract**

**Understanding and tuning correlated states is of great interest and significance to modern condensed matter physics. The recent discovery of unconventional superconductivity and Mott-like insulating states in magic-angle twisted bilayer graphene (tBLG) presents a unique platform to study correlation phenomena, in which the Coulomb energy dominates over the quenched kinetic energy as a result of hybridized flat bands. Extending this approach to the case of twisted multilayer graphene would allow even higher control over the band structure because of the reduced symmetry of the system. Here, we study electronic transport properties in twisted trilayer graphene (tTLG, bilayer on top of monolayer graphene heterostructure). We observed the formation of van Hove singularities which are highly tunable by twist angle and displacement field and can cause strong correlation effects under optimum conditions, including superconducting states. We provide basic theoretical interpretation of the observed electronic structure.**

Van der Waals heterostructures technology provides a variety of tuning knobs, including twist angle, displacement field, and stacking order, for band engineering by precise stacking of one atomically thin crystal onto another[1]. The lattice constant mismatch and relative twist angle give rise to a moiré superlattice, where, under some conditions, interlayer hybridization leads to the formation of an isolated low energy flat band, which quenches the kinetic energy of electronic system. Such low-energy subbands have been realised in several structures and emergent phenomena have been reported, including Mott-like insulators[2], unconventional superconductivity[3-5] and ferromagnetism[6,7] in twisted bilayer graphene (tBLG) and twisted double bilayer graphene (tDBLG)[8-11]. Similar correlated states have also been reported in ABC-trilayer graphene (TLG) superlattice on hexagonal boron nitride (hBN) and rhombohedral stacked graphite films[12-14].

In this work, we study small-angle twisted trilayer graphene (tTLG) van der Waals heterostructures, where a monolayer graphene (MLG) and bilayer graphene (BLG) are stacked and rotated by a small angle with respect to each other. Compared to tBLG, more tuning knobs are expected in tTLG, since the band structures in multi-layer graphene are more tunable than that of the monolayer counterpart[15-18]. In particular, there naturally exists two stacking orders in trilayer graphene, Bernal (ABA)-stacking with mirror symmetry and rhombohedral (ABC)-stacking with inversion symmetry. The former is semimetallic, while the latter is known to be semiconducting with a band gap tunable with an out-of-plane displacement field[19]. The symmetry is further reduced in tTLG, by stacking mono- and bi-layer graphene together with a small relative twist angle[20,21], resulting into ABA-, ABB- and ABC-stacked domains. In this report, we observed electron-hole asymmetry, tunable van Hove singularities as well as correlated insulating states at commensurate fillings on the electron side under finite displacement field in tTLG. The correlated states are asymmetric with respect to *D*, and highly tunable with varying twist angle. In addition, we observed superconductivity signatures in the vicinity of the quarter-filling insulating state. tTLG can be seen as a model system to understand emergent phenomena in the field of twistronics, with twist angle, displacement field and charge density as tuning parameters.

Our twisted graphene heterostructures encapsulated in hBN flakes are fabricated using the recently developed 'tear & stack' method[22,23]. We select the exfoliated graphene flakes with monolayer/bilayer steps to ensure the precise orientation between the two crystals. We use the dual-gate configuration as shown in Fig. 1b, to allow independent tuning of the carrier density *n* and transverse displacement field *D*. By simultaneously applying the top and bottom gate voltages, we obtain $n = \frac{V_{tg}C_{tg}+V_{bg}C_{bg}}{e}$ and $D = \frac{V_{bg}C_{bg}-V_{tg}C_{tg}}{2\varepsilon_0}$, where $C_{tg}$ and $C_{bg}$ are the top and bottom gate capacitances (normalised to unit area) measured from the Hall effect, *e* is the electron charge, and $\varepsilon_0$ is the vacuum permittivity, respectively. The twist angles are determined by the charge density at full filling of each sub-band and the Brown-Zak oscillations[24] (Fig. s3). We have studied more than half a dozen samples with different twist angles ranging between 1.22° and 1.6°. Here we mainly focus on two samples, S1 with twist angle $\theta \approx 1.47°$ and S2 with $\theta \approx 1.22°$.

The schematic of moiré superlattice with a relative twist angle $\theta$ in tTLG, together with the schematics of the transport measurements are presented in Fig. 1a,b. The electronic band structures of mono- and bi-layer graphene hybridize and fold into mini Brillouin zone (MBZ). The size of the moiré unit cell $\lambda$ is given by $\lambda = \frac{a}{2\sin\frac{\theta}{2}}$, where *a*=0.246 nm is the lattice constant of graphene, and the area of moiré unit cell is $A = \frac{\sqrt{3}}{2}\lambda^2$. Each superlattice band in the MBZ can accommodate charge density $n_0 = \frac{4}{A}$, where the prefactor 4 is because of spin and valley degeneracies in graphene and the filling factor is defined as *n/n₀*.

The transport behaviour of our samples is presented in Fig. 1c-g, which display $\rho_{xx}(n/n_0,D)$ maps for samples with $\theta$ from 1.22° to 1.6°, and *n* is normalized to the full-filling charge density $n_0 = \frac{8}{\sqrt{3}\lambda^2}$, associated with the corresponding twist angle $\theta$. Note that tTLG can be fabricated in two mirror-symmetric configurations, either with BLG on top or MLG on top. When we compare the samples with mirror-symmetric configurations, the overall picture flipped (see Fig. s1) which demonstrates that the structure of the states we observe is not an artefact. Therefore, the asymmetry with respect to *D* could be related to the lack of symmetry in tTLG[21]. To be consistent, we define positive *D* when electric fields point from monolayer to bilayer graphene (see Fig. s1).

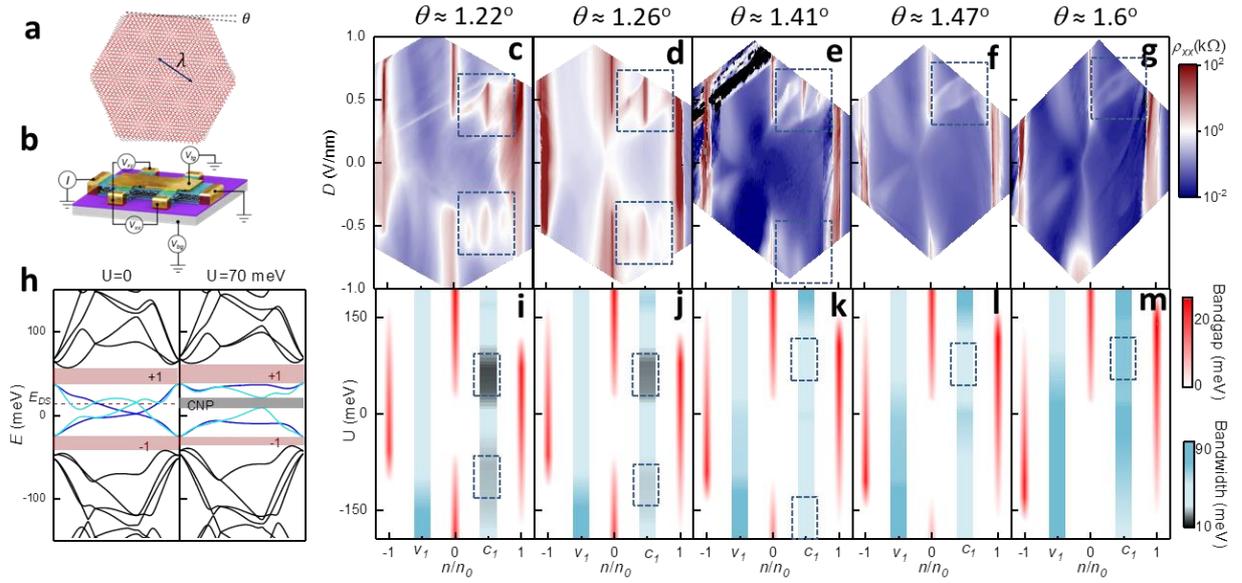

**Fig. 1: Evolution of tTLG band structures with the twist angle. (**a)Illustration of moiré pattern of stacked mono-bilayer graphene with a relative twist angle $\theta$. $\lambda$ is the wavelength of moiré pattern. (b) Schematic of sample structure and measurement configuration. (c-g) $\rho_{xx}(n, D)$ measured at $T$=1.6 K and $B$=0 T of samples with twist angle $\theta \approx$1.22° (c), 1.26° (d), 1.41° (e), 1.47° (f), 1.6° (g). The correlated states under $D$>0 remain almost at the same $D$ range for all samples, while the correlated states under $D$<0 move to larger $D$ when the twist angle increases, and move out of experimentally reachable $D$ in (f) and (g). Black dashed boxes indicate the two correlated regions. (h) Calculated band structure of tTLG with $\theta \approx$1.22° with U=0 (left) and 70 meV (right) and the path in k-space is ($\Gamma \rightarrow M \rightarrow K/K' \rightarrow \Gamma$ ). Red regions indicate the bandgaps at full fillings $\Delta_{\pm 1}$, and grey region shows the gap $\Delta_0$ at CNP opened by U. Cyan (blue) colour indicates the Dirac (parabolic) band inherited from MLG (BLG). Red dashed line indicates $E_{DS}$, the energy of Dirac band shift. (i-m) Calculated single-particle bandgaps and bandwidth of $c_1$ and $v_1$ band as a function of band filling and potential energy difference U for the twist angles studied in experiment. White to red shows the single-particle bangaps, $\Delta_{-1}$, $\Delta_0$ and $\Delta_1$. Grey to blue shows the bandwidth of $c_1$ and $v_1$ bands. Black dashed boxes indicate local minima of bandwidth.

We first discuss the common features emerging in all our devices. First, two resistive peaks emerge at $n = \pm n_0$. Second, at the charge neutrality point (CNP), the resistivity increases with increasing $D$, suggesting a gap opening resulting from the applied displacement field. To understand these features, we calculated the band structures of tTLG with a continuum model[25] (see Supplementary Materials), and Fig. 1h shows an example of such band structure of tTLG with $\theta \approx$1.22°. Cyan (blue) solid line indicates the monolayer (bilayer) graphene band branch after interlayer hybridization, and black indicate higher energy branches. U is the energy potential between the top and bottom layer. When U=0, two single-particle bandgaps emerge at $n/n_0 = \pm 1$ separating the low-energy bands (lowest conduction band $c_1$ and highest valence band $v_1$) from higher energy bands, resulting in the band insulators at full fillings $n/n_0 = \pm 1$. When U=70 meV, the band gap $\Delta_1$ at $n/n_0$=1 increases, while the gap $\Delta_{-1}$ slightly decreases on the hole side. A gap $\Delta_0$ is opened by U at CNP, observed as the increase in the resistivity in experiment. The non-interacting band structure of tTLG is governed by two important features. First, in the twisted system, the Dirac bands are always shifted to positive energy $E_{DS}$>0 even in the absence of any displacement field (see Fig. 1h and explanation below). Second, the shift of the parabolic bands is asymmetric with respect to displacement field (see Fig. s4). Combining these two features i.e. how the bands shift with U, and that $E_{DS} >0$, we can explain the evolution of the band structure with displacement field, and in particular, we expect at low energy: (i) electron-hole asymmetry (in particular, a smaller bandwidth for the conduction band compared to hole band at fixed U); and (ii) asymmetry with the sign of the displacement field (in particular, a smaller bandwidth for conduction band with positive U, compared to negative U).

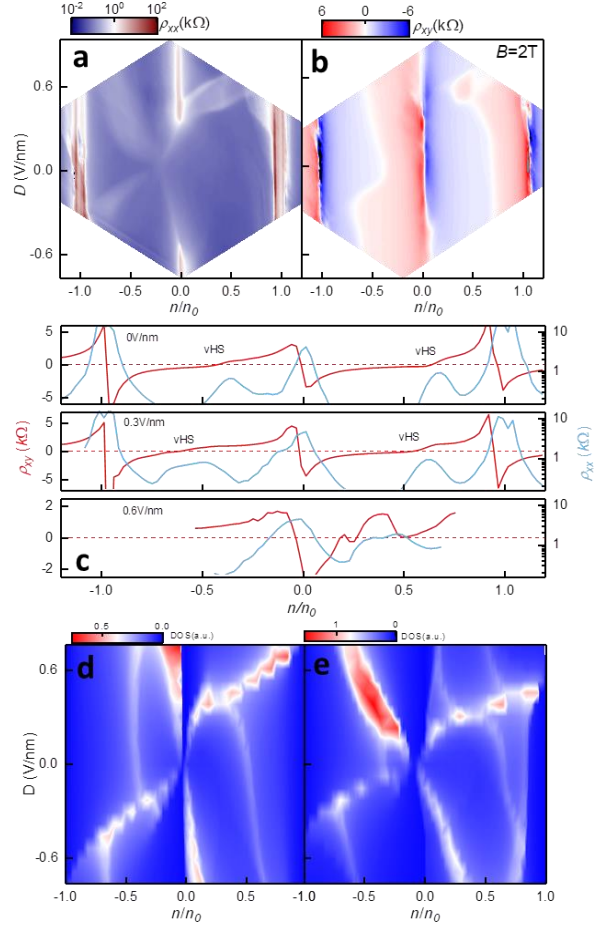

**Fig. 2: Tunable band structure and transport properties of sample S1 with twist angle 1.47°.** (a) $\rho_{xx}(n,D)$ map of S1 measured at $B=0$ T and $T=1.6$ K. Charge density $n$ is normalised to full-filling doping $n_0 = 5 \times 10^{12}$ cm$^{-2}$. (b) $\rho_{xy}(n,D)$ at $B_\perp=2$ T and $T=1.6$ K. (c) $\rho_{xx}$ ($B=0$ T, blue) and $\rho_{xy}$ ($B=2$ T, red) as a function of charge density at $D=0$ (top panel), 0.3 (middle panel) and 0.6 V/nm (bottom panel). Red dashed line indicates $\rho_{xy}=0$. (d-e) Calculated DOS maps for twist angle 1.47° (d) and 1.22° (e) as a function of band filling and displacement field.

In addition we also observe states at half- and quarter-fillings that cannot be explained by the single-particle picture and evolve with twist angle. The corresponding resistivity peaks are strongly asymmetric for electrons and holes, as well as for positive and negative $D$, Fig 1c-g. Similar fractional-filling insulating states have been reported in tBLG[2-4] and tDBLG[8-11], and the states we observe here could be similar correlated states originating from band flatness and band isolation[10]. To confirm the nature of these states, we calculated the bandstructures and extract five features: the bandwidth of $c_1$ and $v_1$ bands, and the single-particle gaps $\Delta_1$, $\Delta_{-1}$, and $\Delta_0$. Fig. 1i-m display the calculated bandgaps and bandwidths and their evolution with twist angle, in which the red stripes at CNP and full fillings show the single-particle bandgaps, and the blue stripes show the bandwidth of $c_1$ and $v_1$ band. In Fig. 1i, two local bandwidth minima are present, as indicated by the black dashed boxes. The bandwidth minimum of $c_1$ band for U>0 increases with $\theta$, consistent with the experimental observation that the correlated feature on the positive *D* side fades away with increasing $\theta$. While for U<0, the position of minimum bandwidth of $c_1$ band moves to larger U. As a result, the flattest band moves out of *D* range that is experimentally achievable for twist angle $\theta$>1.41°, and becomes less isolated from higher conduction bands as $\Delta_1$ decreases with increasing U.

The electron-hole asymmetry could also be explained by the band flatness and band isolation. For example, in Fig. 1i, the band width minimum of $v_1$ band occurs when 80<U<120 meV, where the gap $\Delta_{-1}$ shrinks in the presence of U, meaning the $v_1$ band is less isolated from lower valence bands. With increasing angle, the $v_1$ bandwidth minimum moves to higher U, and the gap $\Delta_{-1}$ is closed. Therefore, the extracted features from our band structures calculation explain the electron-hole asymmetry as well as why the correlated features are asymmetric with *D*, and absent on negative *D* for samples with $\theta$ > 1.41° in the experiment. In fact, the asymmetric features of the band structure is due to the Dirac energy shift[20] where Dirac cone energy states originating from the monolayer are shifted upward compared to the electronic state of the bilayer (Fig. 1h and Fig. s4). We note that the displacement field has two effects on the band structure[20]. First, regardless of the sign of the displacement field, it pushes the conduction band of the monolayer upwards and the valence band downwards with equal magnitude. Second, for parabolic bands originating from the bilayer graphene, a positive U shifts the conduction band upwards by U/2, as might be expected from an isolated BLG, and the valence band downwards by U/6. This effect is reversed for negative U. Notice, therefore, that for positive U, a bandgap at charge neutrality is opened up only after the conduction band of the parabolic bilayer is shifted upwards by the displacement field U≥2$E_{DS}$. The resultant conduction band is very flat because the top of the moiré band is pinned at the *Γ* point. The positive $E_{DS}$ also guarantees that the conduction bandwidth is flatter than the hole bandwidth. For negative U, the band gap at charge neutrality is opened once U≥6$E_{DS}$, and generically the electron and hole bandwidths are larger than for positive U. These features can be clearly seen in the numerical band structures shown in Fig. 1h and Fig. s4, calculated within the continuum model.

Fig. 2a shows the longitudinal resistivity $\rho_{xx}$ of sample S1 as a function of *n* and *D* (*n* is normalized to the full filling-charge density $n_0$= 5 x 10$^{12}$ cm$^{-2}$, associated with a twist angle $\theta$≈1.47°). The angle is confirmed by Brown-Zak oscillation (see Fig. s3). On the hole side, fractional filling resistive peaks emerge in the whole range of *D* measured, and the position of this resistive peak evolves with *D*. While on the electron side, resistive peaks only appear in a small range of *D*, between 0.4-0.7 V/nm, and are absent for *D*<0, in contrast to the case of tDBLG[10], in which the correlated states are symmetric with respect to the sign of *D*.

To characterise the resistive peaks at fractional fillings, we study the response of the sample in a perpendicular magnetic field. Fig. 2b shows the Hall resistivity at *B*=2 T, where sign changes are present at CNP and at full fillings $n = \pm n_0$, in agreement with the presence of single-particle gaps and change of charge type. Unexpectedly, $\rho_{xy}$ also tends to change sign, i.e. Hall density resets, at half filling $n/n_0$=1/2 for 0.4<D<0.7 V/nm in Fig. 2b, which indicates the change of charge type from hole-like to electron-like. Considering the metallic low resistivity (≈200 Ω) in this region, the change of charge type at half filling could indicate the overlap between electron-like and hole-like bands, suggesting new band edge formation as a result of electron correlations when each moiré unit cell hosts 2 electrons.

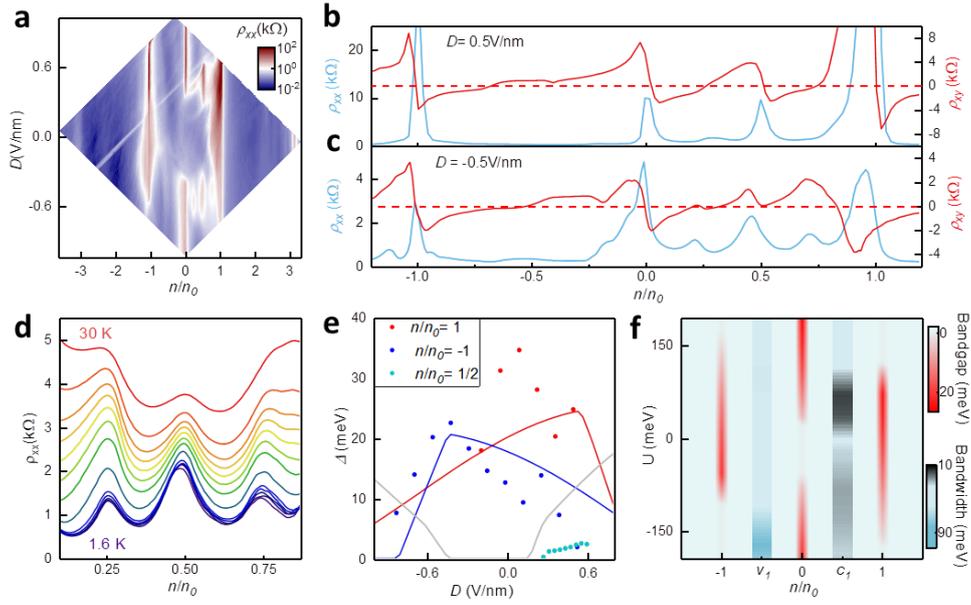

**Fig. 3: Correlated states and transport properties of sample S2 with twist angle 1.22°.** (a) $\rho_{xx}(n,D)$ at $B=0$ T and $T=1.6$ K. Charge density $n$ is normalised to full-filling doping $n_0 = 3.43 \times 10^{12}$ cm$^{-2}$. Correlated states with different manifestations emerge on both positive and negative $D$. (b-c) $\rho_{xx}$($B=0$ T, blue) and $\rho_{xy}$ ($B=2$ T, red) as a function of charge density at $D=0.5$ (b) and $-0.5$ V/nm (c). Red dashed line indicates $\rho_{xy}=0$. (d) Temperature dependence of $\rho_{xx}(n)$ at $D=-0.5$ V/nm. (e) Thermal activation gaps at full-fillings $n/n_0=-1$ (blue circles), $+1$ (red circles) and half-filing (cyan circles). Solid lines are calculated band gaps at CNP (grey), $n/n_0=-1$ (blue) and $+1$ (red). (f) Map of bandgaps and bandwidth as a function of band filling and potential energy difference.

Measurements in low magnetic field allow us to identify the events of the change of band curvature from electron-like to hole-like[26], which are estimated as the states where $\rho_{xy}=0$ (Fig. 2b). Such changes of the band curvature is usually associated with van Hove singularities (vHS). Such vHS appear on both electron- and hole-side, and span the whole $D$ range reachable in experiment. For further clarity, three panels of Fig. 2c display the line cuts of $\rho_{xx}(n/n_0)$ and $\rho_{xy}(n/n_0)$ under $B=2$ T at $D=0$, 0.3 and 0.6 V/nm, respectively. As shown in Fig. 2c, $\rho_{xx}$ peaks emerge at the position of vHS, corresponding to $\rho_{xy} = 0$. Moreover, the position of vHS depends on $D$. On the electron-side, correlated state emerges when the vHS moves to the position at half-filling, Fig. 2b. Therefore, the fractional-filling peaks on the electron side should be related to the vHS based on this observation. Fig. 2d plots the calculated density of states (DOS) map as a function of band filling and displacement field. The positions of DOS peaks in Fig. 2d qualitatively agree with the positon of white stripes in Fig. 2b, confirming the presence of tunable vHS in tTLG.

A much richer phase diagram is observed when the twist angle is close to the *optimal* or so-called *magic angle*. Fig. 3a shows the $\rho_{xx}(n/n_0, D)$ map of sample S2 with twist angle $\theta \approx 1.22°$. Different from sample S1 where resistive peaks ($\approx 200$ Ω) emerge only on positive $D$ between 0.4-0.7 V/nm, sample S2 shows significantly more resistive states (order of 20 kΩ) at quarter- and half-fillings under both positive and negative $D$ (Fig. 3a). To characterize these states, Fig. 3b and Fig. 3c show $\rho_{xx}$ and $\rho_{xy}(B=2$ T) at $D= 0.5$ and $-0.5$ V/nm, respectively. In both Fig. 3b and Fig. 3c, $\rho_{xy}$ changes sign at CNP, and at full fillings $n/n_0=\pm 1$, in agreement with the fact that the Fermi levels pass through single-particle superlattice gaps. The sign changes at $n/n_0=1/2$ at $D=0.5$ V/nm in Fig. 3c indicates the formation of a new Mott-like bandgap as a result of strong correlation. On the contrary, at $D$=-0.5 V/nm, Hall density resettings appear at all commensurate fillings, indicating formations of three new Fermi surfaces when $c_1$ band hosts integer number of electrons. The amplitude of the resistivity for the three states decreases with temperature (after background subtraction, see Fig. 3d, as well as Fig. s2). Therefore, we attribute these three peaks to the correlation at commensurate fillings, and the correlation weakens with increasing $T$. The relatively low resistivity and metallic behaviour of the features at $D$<0 could be because the correlations cause band overlaps, rather than open bandgaps when $c_1$ band hosts integer number of electrons.

The correlated states manifesting themselves as resistive peaks at commensurate fillings $n/n_0$ = 1/4, 1/2, and 3/4, show qualitatively different behaviour for positive and negative $D$. The $D$>0 correlated states are more resistive (order of 20 kΩ) than those on the $D$<0 side (<2.5 kΩ), Fig. 3a. The correlation is much stronger at half filling than at quarter and three-quarter fillings on the positive $D$ side, while on the negative $D$ side, the resistive peaks are of the same order at all commensurate fillings. The correlated states on the positive $D$ side exhibit insulating behaviour, while the negative side are metallic, as shown in Fig. 3d.

In tDBLG[8-10], the correlated states at fractional fillings are attributed to the band flatness and maximum isolation of the first electron $c_1$ band from its neighbouring bands, and in tBLG[2,3], spin and valley degeneracy are broken as a result of strong correlation, giving rise to the correlated states at integer number of electron/hole fillings. Fig. 3f shows the bandwidth of $c_1$ and $v_1$ bands, as well as the bandgaps at CNP and $n/n_0$=±1. Indeed, we find flattest $c_1$ band occur at positive potential difference 40<U<70 meV, where both the gaps at CNP and $n/n_0$=1 are present. In addition, a local minimum bandwidth also emerges for -120<U<-70 meV. The observation that the $c_1$ band is flatter when U>0 than that when U<0 agrees with the experimental observation that correlated states under positive $D$ are more robust than that under negative $D$. Fig. 3e displays the experimental thermal activation gaps at $n/n_0$=±1, and the calculated single-particle gaps as a function of $D$, which qualitatively agree well with each other, confirming that the band isolation plays a role in the correlation.

For a correlation mediated insulator, being close to commensuration is necessary but not sufficient condition. Similar to tBLG, the origin of the strongly correlated insulator is not yet understood. It could arise from the formation of a Mott insulator[2], a Wigner crystal[27], or because the electron interactions lift the spin or valley degrees of freedom[5]. For all of these mechanisms, the insulating state is expected to occur at commensurate fillings. This is because electron-electron interactions conserve momentum, and can only dissipate current by the Umklapp processes that are enabled when a moiré subband is completely filled[28].

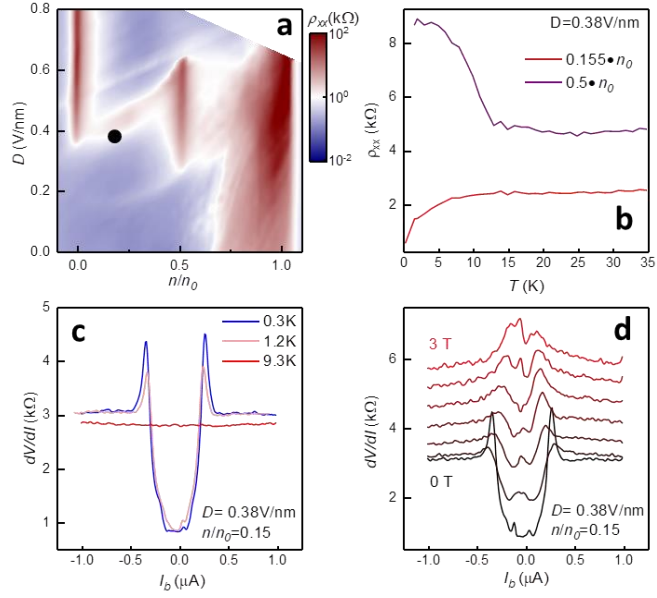

**Fig 4: Signatures of superconductivity in tTLG near electron quarter filling correlated state.** (a) Zoom-in of Fig. 3a between CNP and electron full filling. Black circle shows the region where superconductivity is observed. (b) Resistivity versus temperature under $D$=0.38 V/nm. The insulating response at half filling (red) onsets ≈12K, and the resistivity near superconducting regime decreases sharply from ≈2.5 kΩ to ≈ 400 Ω at 0.3 K (black). (c) Response of $dV/dI$ at superconducting state $D$=0.38 V/nm, $n/n_0$=0.15 as a function of DC current $I_b$ at 9.3 K (red), 1.2 K (pink) and 0.3 K (blue). (d) dV/dI at superconducting state ($D$=0.38 V/nm, $n/n_0$=0.15) as a function of perpendicular $B$ field from 0 T to 3 T, increasing by a step of 0.5 T.

Since it is anticipated both experimentally and theoretically[3,29-31] that superconductivity occurs for doping away from commensurate filling, we look for superconductivity in our tTLG experiment close to the commensurate filling $n/n_0$=1/4, and in regions where the non-interacting band structure has a peak in the DOS (Fig. 2e). Fig. 4 displays the signature of superconductivity near electron quarter-filling at positive $D$=0.38 V/nm in sample S2. A small excitation current 1nA is used to measure the resistance. Fig. 4b shows temperature dependence of resistivity at half filling $n/n_0$=1/2 and at region near quarter filling $n/n_0$=0.155 at $D$=0.38 V/nm. Half-filling state displays insulating behaviour, while the resistivity at doping $n$=0.155$n_0$ drops abruptly at low temperatures. To verify its superconducting nature, we measure differential resistance $dV/dI$ as a function of driving DC current in varying temperatures as shown in Fig. 4c. At low temperature, a typical superconducting $dV/dI(I_b)$ is observed, where two sharp peaks are present to define the critical current $I_c$≈260 nA at 0.3 K, excluding the Joule heating mechanism of the non-linear $I$-$V$ curves[32]. Increasing temperature drives the sample into normal metallic state. Perpendicular magnetic field also suppresses the superconductivity as shown in Fig. 4d. The observations confirm the emergence of superconductivity near quarter-filling under $D$=0.38 V/nm. The non-zero resistivity in the superconductivity regime could be attributed to sample inhomogeneity or non-ideal contacts, similar to previous reports[4,13].

In summary, tTLG is a perfect model to study tunable vHS and emergent quantum phenomena resulting from electronic correlations in engineered flat bands. We demonstrated that correlated states in tTLG can be turned on and off by a displacement field. Moreover, the reduced symmetry in tTLG makes the correlated states highly tunable by twist angle, and two asymmetric correlated regions with respect to *D* are observed in our samples. In addition to the interesting correlation physics, the Bernal (ABA) and rhombohedral (ABC) stacked domains in tTLG provide ideal platforms to study topological edge states as ABA- and ABC-stacked few-layer graphene host totally different band structures[33]. Our work thus paves the way for understanding the mechanism of strong correlations in twistronics.

## Methods

The heterostructures of twisted trilayer graphene encapsulated by hBN were assembled by the standard dry-transfer and tear-and-stack techniques[22] with a polypropylene carbonate (PPC) film on top of a polydimethylsiloxane (PDMS) stamp. Graphene flake with monolayer/bilayer steps and hBN flakes were exfoliated onto $SiO_2$/Si substrate. After picking up a top hBN flake with the PPC/PDMS stamp, we used the stack to tear and pick up the monolayer part of graphene flake from the bilayer part at the temperature of 40 ˚C. The remaining bilayer part on the substrate was then rotated by a desired angle and subsequently picked up by the monolayer graphene/hBN stack. The heterostructures were then released onto a bottom hBN exfoliated on 300 nm $SiO_2$/doped Si substrate to form hBN/monolayer Gr/bilayer Gr/hBN stack. The standard Hall bar geometry of the devices was defined by e-beam lithography and etched by $CHF_3/O_2$ plasma. The one-dimensional electrical contacts and top gate were deposited by e-beam evaporation of Cr/Au.

Transport measurements were performed in an either dry or liquid helium cryostat using standard low-frequency AC measurement with Lock-in amplifiers. To detect the superconductivity, the sample was loaded into $^3$He cryostat. Low temperature electrical filters were used to increase the signal-to-noise ratio. The difference resistance dV/dI were measured by coupling a small AC current (1 nA) and a DC current bias and measuring the differential voltage using lock-in technique.

# Supplementary Materials

1. Electronic properties of twisted trilayer graphene

Among the different van der Waals materials that can be twisted on each other, our system is rather unique. For hetero-bilayers like graphene twisted on hexagonal boron nitride (hBN), the lattice mismatch between carbon lattice and the BN lattice is the dominant physics, as is the differential coupling between boron and nitrogen with carbon[34]. Since our system is all comprised of carbon layers, we do not have any such complications of hetero bilayers. Yet, MLG and BLG are electronically distinct materials. MLG has massless fermions with π Berry's phase, while BLG has massive fermions with 2π Berry's phase. The moiré bands formed by hybridizing these have weight on all three layers, and inherit the properties of both systems[35]. We know that MLG is much more susceptible to disorder compared to BLG[36], while BLG require much larger magnetic fields to have the cyclotron energy comparable to the kinetic energy. Our system combines the advantages of both. Also, an isolated MLG preserves electron-hole symmetry, and is not susceptible to a displacement field, while an isolated BLG breaks both. In our system, these are only weakly broken by higher order processes that couple the two systems, thereby giving us the advantages of both.

2. Continuum model of twisted trilayer graphene

To theoretically study the system of twisted monolayer-bilayer graphene, we built a continuum model following the approach introduced in Ref[25]. In our case we have a monolayer coupled to a bilayer graphene by three 2x4 tunneling matrices in contrast to the 2x2 matrices used to couple two monolayer graphene.

The primitive lattice vectors for monolayer and bilayer graphene are defined as

$$a_1 = \left(\frac{a}{2}, \frac{\sqrt{3}\,a}{2}\right) \quad , \quad a_2 = \left(\frac{a}{2}, \frac{-\sqrt{3}\,a}{2}\right)$$

where a= 0.246 nm is the lattice constant of graphene.

The displacement vectors from an atom A to the nearest three B atoms are

$$\delta_1 = \left(0, \frac{a}{\sqrt{3}}\right), \quad \delta_2 = \left(\frac{a}{2}, \frac{-a}{2\sqrt{3}}\right), \quad \delta_3 = \left(\frac{-a}{2}, \frac{-a}{2\sqrt{3}}\right)$$

The primitive reciprocal lattice vectors are given by

$$b_1 = \left(\frac{2\pi}{a}, \frac{2\pi}{\sqrt{3}\,a}\right), \quad b_2 = \left(\frac{2\pi}{a}, \frac{-2\pi}{\sqrt{3}\,a}\right)$$

Considering $p_z$ orbitals on $A_1$ and $B_1$ sublattices, the tight-binding description for monolayer graphene is

$$H_{MLG} = \begin{pmatrix} -\frac{U}{2} & \gamma_0 f(k) \\ \gamma_0 f^*(k) & -\frac{U}{2} \end{pmatrix}$$

where f(k) describes the nearest neighbor hopping, given by

$$f(k) = \sum_{j=1}^{3} e^{i k.\delta_j}$$

In the tight-binding description for AB-stacked bilayer graphene, we take four $p_z$ orbitals on $A_2$, $B_2$, $A_3$ and $B_3$ sites. The resulting Hamiltonian is

$$H_{BLG} = \begin{pmatrix} 0 & \gamma_0 f(k) & \gamma_4 f(k) & \gamma_3 f^*(k) \\ \gamma_0 f^*(k) & 0 & \gamma_1 & \gamma_4 f(k) \\ \gamma_4 f^*(k) & \gamma_1 & \frac{U}{2} & \gamma_0 f(k) \\ \gamma_3 f(k) & \gamma_4 f^*(k) & \gamma_0 f^*(k) & \frac{U}{2} \end{pmatrix}$$

with parameter values determined by DFT[37] as

$$(\gamma_0, \gamma_1, \gamma_3, \gamma_4) = (-2.61,\ 0.361,\ 0.283,\ 0.138)\ eV$$

Layer 1 and Layer 2 are coupled by three 2x 4 tunneling matrices

$$T_j = \begin{pmatrix} w_{AA} & e^{-i(j-1)\phi} w_{AB} & 0 & 0 \\ e^{i(j-1)\phi} w_{AB} & w_{AA} & 0 & 0 \end{pmatrix}$$

where $\phi = \frac{2\pi}{3}$ and j=1,2,3, $w_{AA}$= 0.050 eV is the Fourier component of the tunneling between sublattice $A_1$ and sublattice $A_2$ and $w_{AB}$= 0.085 eV is the Fourier component of the tunneling between sublattice $A_1$ and sublattice $B_2$[38].

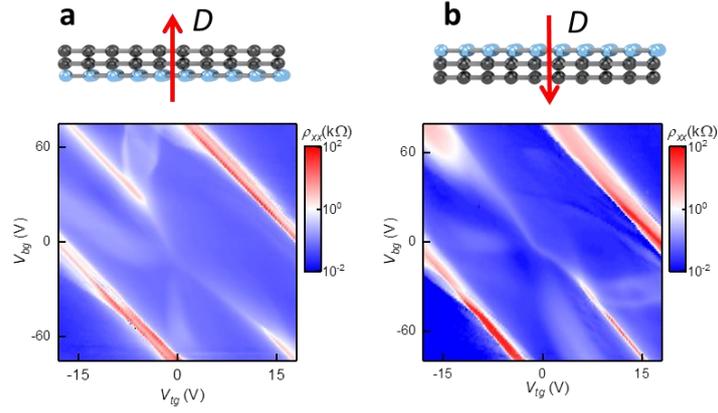

**Fig. s1: Two mirror-symmetric configurations of tTLG.** $\rho_{xx}$ maps as a function of back gate voltage $V_{bg}$ and top gate voltage $V_{tg}$ of samples with twist angle $\theta \approx 1.47°$ (a) and $1.6°$ (b) measured at $T=1.6$ K and $B=0$ T. The schematics above the maps show the stacking configuration. Black and blue balls indicate bilayer and monolayer graphene, respectively. Red arrows define positive $D$.

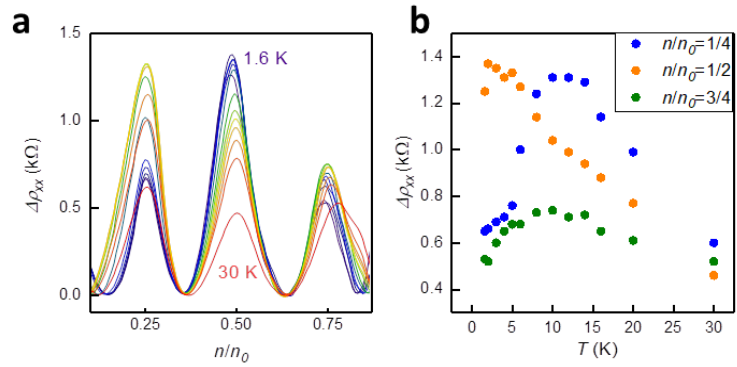

**Fig. s2: Temperature dependence of correlated states under D<0 in sample S2.** (a) The same figure as Fig. 3d in the main text, but after subtracting a smooth background. (b) Peak amplitudes as a function of temperature at fractional electron fillings.

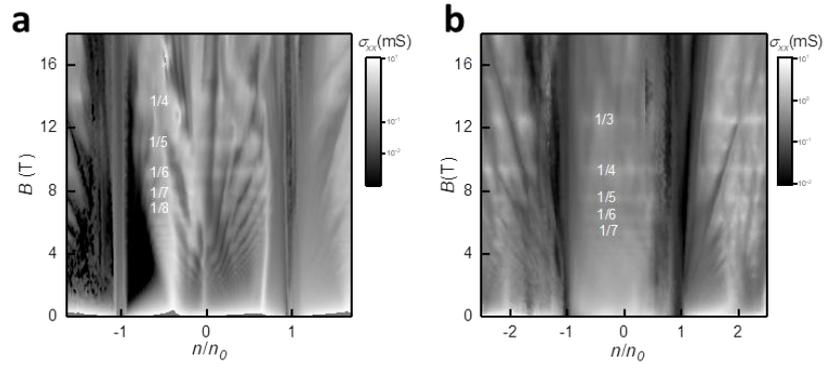

**Fig. s3: Brown-Zak oscillations.** $\sigma_{xx}$ maps of sample S1 (a) and S2 (b) versus normalised charge carrier density and magnetic field measured at $T$=1.6K and $D$=0. Numbers indicate $\frac{\Phi_0}{\Phi}$ fractions, where $\Phi_0 = \frac{h}{e}$, $h$ is Planck constant, $e$ is electron charge, and $\Phi = BS$ is magnetic flux through a moiré unit cell.

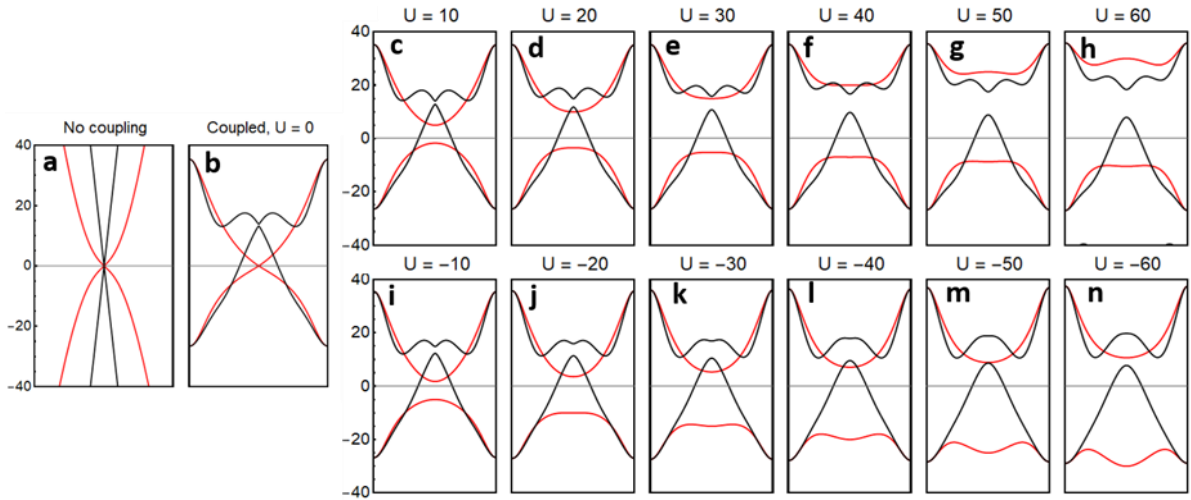

**Fig. s4: Dirac energy shift in the asymmetric evolution of band structure in tTLG.** Energy units are in meV and the path in k-space is ($\Gamma \rightarrow K/K' \rightarrow \Gamma$). (a) Band structure without coupling the two subsystems through the tunnelling matrices $T$. (b) Band structure with non-zero tunnelling matrices for zero interlayer potential energy difference U=0. The coupling of the two subsystems even without applying potential difference U results in shifting the Dirac cone upwards. (c-h) Effect of applying positive interlayer potential energy difference U>0. The bandgap at CNP needs U smaller than 30 meV to open. Moreover, electron band $c_1$ is flatter than the hole band $v_1$. (i-n) Effect of applying negative interlayer potential energy difference U<0. In contrast to when U>0, it requires larger energy difference U< -50 meV to open a gap at CNP.

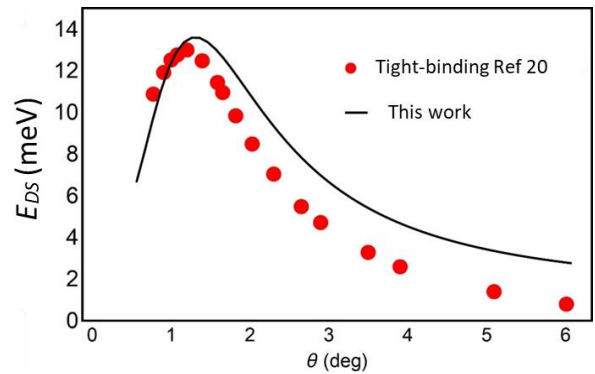

**Fig. s5: Dirac energy shift as a function of twist angle.** Solid black curve is the Dirac energy shift produced by our continuum model and the red dots are the results of tight-binding model in Ref [20].